\documentclass[useAMS,usenatbib]{mn2e}
\pdfoutput=1
\usepackage{amsmath}
\usepackage{amssymb}
\usepackage{microtype}
\usepackage{mathptmx}
\usepackage{graphicx}
\usepackage{booktabs}
\usepackage{aas-macros}
\usepackage{epstopdf}

\newcommand{\HII}{H\,{\sc ii}}

\newcommand{\unit}[1]{\mathrm{#1}}
\newcommand{\unitp}[2]{\ensuremath{\mathrm{#1}^{#2}}}
\newcommand{\MSolar}{\ensuremath{\textrm{M}_{\odot}}}

\newcommand{\Halpha}{\ensuremath{\mathrm{H}{\alpha}}}
\def\micron{\hbox{$\mu$m}}

\title[Observations of M\,66 at 90\,GHz]{90\,GHz Continuum Observations of Messier\,66 }
\author[B. Nikolic and R.~C. Bolton]{B. Nikolic\thanks{E-mail:
    b.nikolic@mrao.cam.ac.uk} and R. C. Bolton\\Astrophysics Group, Cavendish Laboratory, University of Cambridge,
Cambridge CB3 0HE, UK}

\begin{document}

\date{DRAFT}

\pagerange{\pageref{firstpage}--\pageref{lastpage}} \pubyear{2011}

\maketitle

\label{firstpage}

\begin{abstract}
  Radio emission at around 90\,GHz from star-forming galaxies is
  expected to be strongly dominated by the free-free component due to
  ionising radiation from massive, short-lived, stars. We present high
  surface-brightness sensitivity observations at 90\,GHz of the nearby
  star-forming galaxy Messier 66 with resolution of about 9\,arcsec
  (corresponding to a physical scale of about 500\,pc) and analyse
  these observations in combination with archival lower
  frequency radio and mid-infrared measurements.  For the
    four regions for which the observations support our models we find
    that the free-free component indeed dominates the emission at
    90\,GHz, making up 76--90 per cent of the luminosity at this
    frequency but with the data also consistent with all of the
    emission being due to free-free.  The estimates of free-free
    luminosities are also consistent, within measurement
    and decomposition errors, with star-formation rates derived from
  lower radio frequencies and mid-infrared observations. In our
  analysis we consider both power-law and curved spectra for the
  synchrotron component but do not find evidence to support the curved
  model in preference to the power-law.
\end{abstract}


\section{Introduction}

Observations of continuum emission at radio frequencies can provide
useful insight into star-formation activity in external galaxies
\citep[see, for example, the review by][]{1992ARA&A..30..575C}. This
is greatly aided by the fact that radio frequency radiation is 
unaffected by dust absorption, and by the use of interferometer arrays 
to give high angular resolution and astrometric accuracy even at long wavelengths.

The continuum emission at these frequencies is due to a combination of
two mechanisms that each trace different, but important, stages of
evolution of recently formed and short-lived, high-mass stars
\citep{1992ARA&A..30..575C}:
    \begin{enumerate}
    \item Free-free radiation (presumably dominant at higher
      frequencies) which traces the ionising ultraviolet continuum photons
      from massive young stars whilst they are on the main sequence. 
    \item Synchrotron radiation, which is dominant at lower radio frequencies
    and which traces relativistic electrons likely created by supernovae from the same massive 
   stars at the end of their lives.
\end{enumerate}

This utility of radio observations for studying star-formation in
external galaxies is already well established at lower frequencies,
for example, through large-area, sensitive surveys at frequencies
around 1.4\,GHz, such as the FIRST \citep{1995ApJ...450..559B} and
NVSS \citep{1998AJ....115.1693C} surveys. In particular, the observed
tight correlation between radio and far-infrared luminosities of local
\citep[e.g.][]{1985ApJ...298L...7H} and more distant
\citep[e.g.][]{2010MNRAS.409...92J} galaxies provides important
empirical evidence for use of radio-continuum as a tracer of recent
star formation activity.

Observations at higher frequencies ($\nu \gtrsim 10\,\unit{GHz}$) are
more challenging than at frequencies around 1.4\,GHz because the
surface brightness of emission from galaxies is fainter while the
receiver noise and atmospheric effects are greater. Such observations
can, nevertheless, be extremely useful.

High frequency measurements, where total flux density is dominated 
by free-free emission should enable astronomers to make accurate 
estimates of both the synchrotron and the free-free components of 
the emission when combined with data at lower frequencies. The 
contribution from the free-free component has a a direct physical 
link to emission in \Halpha\ and other Hydrogen recombination lines 
which are some of the most commonly used tracers of star-formation. 
At the same time though, radio measurements retain the advantage 
of being unaffected by dust extinction. Hence accurate estimates of 
free-free emission can be used to calibrate \Halpha\ other measures 
of star-formation in dusty environments \citep[see for example][]
{2011ApJ...737...67M}.

Extraction of the flux density of the synchrotron component
places constraints on the \emph{shape\/} of the synchrotron
spectrum. At frequencies above 1\,GHz, this shape is determined by the
energy distribution of relativistic electrons in the source, which in
turn is determined by the energy losses of the electrons accelerated
in the supernova remnants.  Therefore the shape of the synchrotron spectrum
can constrain the evolution of recent star formation activity
or cooling processes of the relativistic electrons.

Recently, the combination of large single-dish telescopes and
broad-band multi-pixel receivers has significantly improved the
achievable surface brightness sensitivity of radio continuum
observations at frequencies between 30--100\,GHz, while giving angular
resolution well matched to characteristic physical length-scales of
star-formation in nearby galaxies.  We have therefore begun a
programme of observations aimed at characterising the free-free
emission from nearby galaxies by spatially resolved mapping
observations at 90\,GHz.  Our pilot observations were made at the
Green Bank Robert C. Byrd 100-m diameter Telescope \citep[GBT,
see][]{2004SPIE.5489..312J} using the MUSTANG\footnote{{\bf
    MU}ltiplexed {\bf S}QUID {\bf T}ES {\bf A}rray at {\bf N}inety
  {\bf G}Hz} 90\,GHz 64-pixel bolometer array with 20\,GHz bandwidth
\citep{2008SPIE.7020E...4D}.

For this initial study it was clear that we needed a comprehensive set
of multi-wavelength data so targets were selected from the well
observed {\it Spitzer\/} Infrared Nearby Galaxies Survey (SINGS)
\citep{2003PASP..115..928K} sample. We selected two targets: Messier
66 (M\,66, NGC 3627) and Messier 99 (M\,99, NGC 4254), which are the
two galaxies in the SINGS with highest star-formation rates as given
by \cite{2003PASP..115..928K}. Both of these galaxies were observed
but did we not reliably detect M\,99, hence in this paper show only
the results for M\,66.

\section{Observations and data processing}

\subsection{GBT/MUSTANG observations}

\begin{figure}
  \begin{centering}

    \includegraphics[width=1.20\columnwidth,angle=-90]{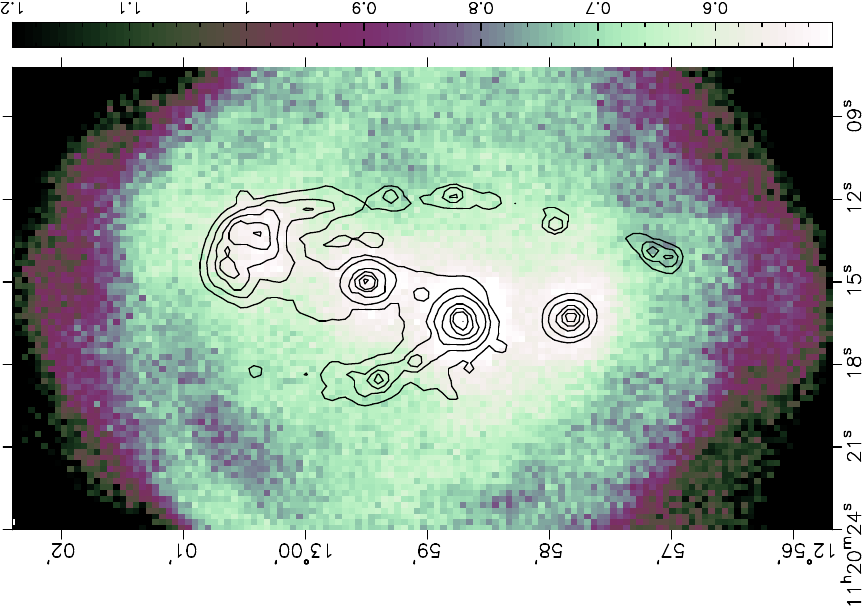}

  \end{centering}
  \caption{The estimated noise in the final map produced from the
    GBT/MUSTANG observations shown as colour scale (units are
    mJy/Beam).  Also shown as contours is the 24\,\micron\
    \emph{Spitzer\/} map of this galaxy which was used to decide where
    to target the telescope time. The contours are drawn at
    $(1/2)^n\times \textrm{maximum}$.}
\label{fig:noise-24mi}
\end{figure}

\begin{figure}

  \begin{centering}
    \includegraphics[width=1.2\columnwidth,angle=-90]{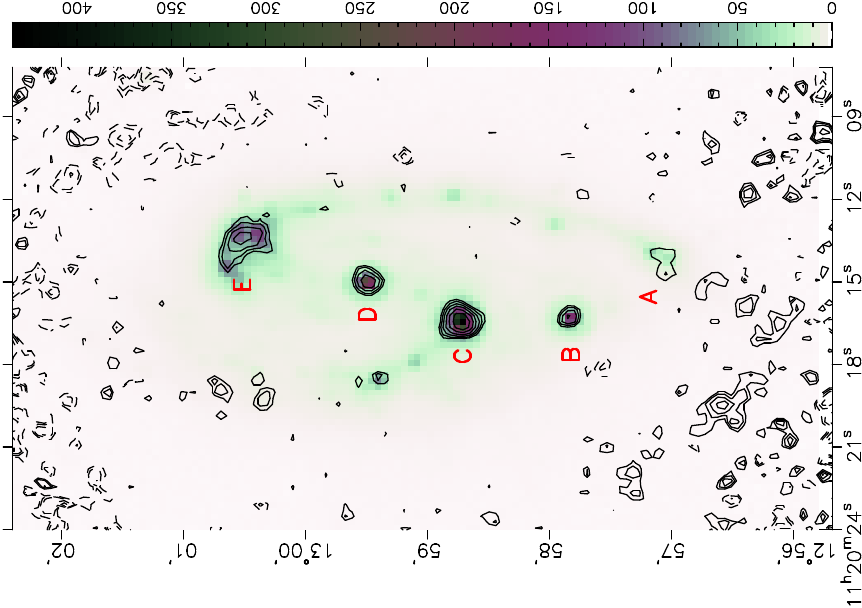}
  \end{centering}

  \caption{Contours of the results of our GBT+MUSTANG observations
    overlaid on a \emph{Spitzer} $24\,\micron$ image. The contours are
    drawn at intervals of $(3/4)^n\times \textrm{maximum}$ and are at
    1.02, 0.76, 0.57, 0.43, 0.32, 0.24 and
    0.18$\times$\,MJy\,\unitp{sr}{-1}.  The colour scale (representing
    the $24\,\micron$ emission) is also in units of
    MJy\,\unitp{sr}{-1}. For better presentation the MUSTANG map has
    been smoothed by a 6\,arcsec Gaussian filter before plotting of
    the contours giving resolution in this diagram of 11.3\,arcsec. }
  \label{fig:mustang-on-24}

\end{figure}

Observations with GBT/MUSTANG were made in the winter 2010/2011
observing season under projects GBT09C-020 and GBT10C-011. As the
field of view of MUSTANG (about $40\times40\,$arcsec) is much
smaller than the angular extent of M\,66 (about $5\times3$ arcminutes)
the observation required the use of a scanning strategy. The choice of
the scanning strategy involves a trade-off between area covered and
the sensitivity and accuracy of measured fluxes. We decided on a
combination of two strategies:
\begin{enumerate}
\item A `daisy-petal' strategy which concentrates a large fraction of
  the integration time in a small angular region. We split the time
  approximately equally between four centres that were selected based
  on the \emph{Sptizer\/} 24\,\micron\ image of the galaxy;
\item A billiard-ball box scanning strategy which distributes
  integration approximately uniformly over a rectangular area on the
  sky. We used box size that was 3 arcminutes on the side and we used
  three slightly offset centres for the box scanning pattern.
\end{enumerate}

Our total on-source observing time of 5.5 hours was split between
these two strategies and all the data from both strategies were
analysed together. The resulting noise map is shown in
Figure~\ref{fig:noise-24mi} with contours of \emph{Sptizer\/}
24\,\micron\ image overlaid on top. It can be seen that although the
sensitivity of our observations is best around the prominent knots of
24\,micron\ emission that we were targeting, there is also good
sensitivity over a wide, approximately square, area that encompasses
the main morphological features of the inner part of this galaxy.

At the beginning at each observing block, and occasionally at other
times, the surface of the GBT was adjusted using the Out-Of-Focus
(OOF) holography technique \citep{2007A&A...465..685N} in order to
correct for the thermal deformations of the telescope structure. This
procedure improves the overall efficiency of observations and also
improves the shape of the primary beam of the telescope. During each
session the flux calibration standard $\alpha\,$Orianis was observed.
Secondary calibrators were also observed, about once per hour, to
check the beam shape and enable off-line correction for
  pointing variation. Our pointing calibrator maps show a beam size of
  9.6\,arcsec.  The observed data were reduced using the suite of
IDL-based routines provided by the MUSTANG instrument
team. The data reduction strategy and algorithms implemented
  in this software are described by
  \cite{2010ApJ...716..739M}. We also produced a MUSTANG map
  smoothed to 15\,arcsec resolution for photometry with matched
  resolution to lower frequency radio data as described below.

\subsection{Archival observations}

We complement our GBT/MUSTANG observations with archival radio,
\Halpha\ and mid-IR observations. We used the NRAO VLA
archive\footnote{\url{https://archive.nrao.edu/archive/}} to obtain
pre-processed images of M\,66 at L, C and X bands (1.4, 4.8 and
$8.5\,\unit{GHz}$ respectively). 

The largest angular scales of emission that the VLA observations are
sensitive to are 15--3 arcminutes, which is significantly larger than
the largest angular scale that the MUSTANG observations are sensitive
to (about 40\,arcsec, as discussed below).

At L-band we used data from project AS541, observed by
\cite{2009A&A...503..747P}. Originally this map had a beam of
$14.9\times14.3$\,arcsec: this was smoothed to
$15.0\times15.0\,$arcsec (for ease at other
  frequencies).  We use C and X band data from more than one
observation.  The different maps in each band were first smoothed to
$15\times15$\,arcsec resolution before being reprojected (using the
CASA task \texttt{regrid}) onto the same basis and then combined via a
variance-weighted sum (to give the best off-source noise in the final
map).

The final C band map uses data from VLA projects AU078 and AS0551,
which had initial resolutions of $\sim13$ and $\sim14\,\unit{arcsec}$,
respectively, prior to smoothing to $15\,\unit{arcsec}$.  The final X
band map uses data from VLA projects AS0551, AU075 and AU078.  These
had initial resolutions of $\sim\unit{9}$, $\sim7.3$ and
$\sim7.6\,$arcsec respectively, prior to smoothing to
$15\,$arcsec.

Our 24\,\micron\ and \Halpha\ data are derived from the
  enhanced data products released by the SINGS collaboration
  \citep{2003PASP..115..928K}. We used the data released on 10 April
  2007 (DR5) as available from IPAC web-site{\footnote{
      \url{http://irsa.ipac.caltech.edu/data/SPITZER/docs/spitzermission/observingprograms/legacy/sings/}}}. For
  estimation of \Halpha\ luminosities we use the continuum subtracted
  image which reduces the contamination due to stellar continuum from
  within M\,66 and any faint foreground stars within the measurement
  apertures (no bright foreground stars are obvious within the
  apertures). The photometric uncertainty of \Halpha\ images before
  continuum subtraction is 10 per-cent so after subtraction the
  uncertainty is likely to be of order of 20 per-cent. We did not
  apply a correction for extinction of \Halpha\ light within the Milky
  Way (approximately 0.1 magnitude according to maps of
  \citealt{1998ApJ...500..525S}) as the intrinsic extinction within
  M\,66 is likely to be much higher.  We also produced versions of
  24\,\micron\ and \Halpha\ maps smoothed to resolution of 15\,arcsec
  which we used for photometry at resolution which was matched to both
  radio and MUSTANG data.

\section{Analysis and Results}

\begin{table*}
\begin{tabular}{ccccl}
\toprule
Label & RA & Dec & Radius & Notes \\
      & (J2000)   &  (J2000)    & (arcsec)&    \\
\midrule
A    & 11:20:14.0 & 12:57:05.8 &  12.5 & Southern-most part of western
spiral arm\\
B    & 11:20:16.3 & 12:57:48.9 &   13.7& High-recessional velocity, tucked in part of western arm?\\
C    & 11:20:16.4& 12:58:42.2 &  16.1  & Southern bar end\\
D    & 11:20:15.0& 12:59:29.0 &  14.4  & Galaxy nucleus  \\
E    & 11:20:13.7& 13:00:30.5 &  25.9$\times$16.5 & Northern bar end \\
\bottomrule
\end{tabular}

\caption{Coordinates and sizes of regions used in the analysis of
  90\,GHz luminosity. The sizes are expressed as radii for regions
  A--D (which all circular) and as semi-major$\times$ semi-minor axis
  for region E (which is elliptical with the major axis
  parallel to the equator).} 
\label{tab:regions}

\end{table*}

We analyse the MUSTANG observations quantitatively by identifying
regions of significant emissions and extracting fluxes for these
regions from maps made at lower radio frequencies and from MUSTANG
observations. Because of the data processing required to remove the
rapidly varying atmospheric effects, our MUSTANG observations have
limited sensitivity to emission on angular scales larger than about
40\,arcsec (see the recent work by Mason et
al\footnote{\url{https://safe.nrao.edu/wiki/bin/view/GB/Pennarray/NewPipelineCharacterization}}). As
a result our selected regions concentrate on knots of emission smaller
than 40\,arcsec and it is not possible to attempt an analysis of
the total flux from this galaxy.

To identify regions of significant emission we smoothed the MUSTANG
map to a resolution of 15\,arcsec (which is the resolution of our
lower-frequency radio data) and identified by eye in this map five
regions of well-detected 90\,GHz emission. The regions are labelled on
Figure~\ref{fig:mustang-on-24} and their coordinates and sizes are
shown in Table~\ref{tab:regions}.  We carried out photometry by simple
addition of flux densities in the regions; for
  `A'--`D' we used circular apertures with centres and radii as given
  in Table~\ref{tab:regions} and for region `E' we used an elliptical
  aperture with the centre, semi-major and semi-minor axis also given
  in Table ~\ref{tab:regions} and with the major axis parallel to the
  equator.  All of the regions are signifantly larger than
  the resolution of the observed maps and in the case of {\it
    Spitzer\/} they also contain the first diffraction ring of the
  telescope at 24\,\micron. Nevertheless, to reduce the effects of
  different resolutions, all of the photometry measurements were made
  on maps smoothed to 15\,arcsec resolution, set by our lowest
  resolution radio-frequency data set.

We did not attempt to calculate and subtract local diffuse
background. As described above, MUSTANG data processing regardless
filters out most emission on angular scales larger than about
40\,arcsec, effectively removing the diffuse emission. Lower
  frequency radio observations were made with minimum spacing of
  interferometer elements such that much larger scales are
  retained. Therefore, at lower frequencies there is some additional
  uncertainty in flux density measurements due to this
  diffuse local background. As the source of this local background may
  be diffusion of relativistic electrons from knots of star-formation,
  any attempt to remove it may make interpretation of subsequent
  results much more difficult and so we have chosen not to do so. The
  24\,\micron\ and \Halpha\ are `total power' measurements, i.e., all
  angular scales larger than the angular resolution are retained, and
  so these data also have additional uncertainty associated with
  diffuse local background. \cite{2011ApJ...737...67M} find better
  match between estimates of star-formation rates from synchrotron and
  free-free components when the local background is not removed and
  also find that background removal does not significantly affect the
  decomposition of spectra into free-free and synchrotron components.

For the following analysis we adopt a distance to M\,66 of 11.1\,Mpc
derived from Cepheid measurement by \cite{1999ApJ...522..802S}, which
means that one arcminute corresponds to approximately
$3.2\,\unit{kpc}$ projected distance.

\begin{table*}
\begin{tabular}{ccccccccl}
\toprule
Label & $F_{\nu}(1.4\,\unit{GHz})$ & $F_{\nu}(4.9\,\unit{GHz})$&$F_{\nu}(8.5\,\unit{GHz})$ & $F_{\nu}(90\,\unit{GHz})$ & $F_{\nu}(24\,\micron)$\\
      & (mJy)                  &   (mJy)                    & (mJy)
      &(mJy) & (mJy)    \\
\midrule
A  & $1.9\pm0.1$ & $1.5\pm0.1$ & $0.9\pm0.05$& $1.6\pm0.4$&     110  \\  
B  & $9.1\pm0.5$ & $4.6\pm0.2$ & $2.8\pm0.1$ & $1.7\pm0.3$&     310  \\
C  & $39.6\pm2.0$  & $18.3\pm1$  & $12.0\pm0.6$& $7.4\pm1.1$&   1160 \\
D  & $24.1\pm1.0$  & $10.6\pm0.6$ & $6.0\pm0.3$& $2.7\pm0.4$&   490  \\
E  & $47.6\pm2.0$  & $22.6\pm1.1$ & $14.4\pm0.7$& $7.4\pm1.1$ & 1190 \\
\bottomrule
\end{tabular}

\caption{Measured fluxes densities of regions as describe in Table~\ref{tab:regions}.} 
\label{tab:fluxestable}

\end{table*}

\begin{table*}
\begin{tabular}{ccccccccl}
\toprule
Region& $\left<\nu_{\rm SN}\right>$ & $\left<\alpha\right>$ & $\left<\Psi\right>$                & $\Psi_{\Halpha}$ & $\Psi_{24\,\micron}$ & $f_{{\rm th},90\,\unit{GHz}}$ & $f_{{\rm th},33\,\unit{GHz}}$\\
      & ($10^{-3}\unitp{yr}{-1}$)  &                       & ($\unit{\MSolar}\,\unitp{yr}{-1}$)  & ($\unit{\MSolar}\,\unitp{yr}{-1}$) & ($\unit{\MSolar}\,\unitp{yr}{-1}$)\\
\midrule
A  &        &       &     $0.20\pm0.07^{*}$          & 0.024 & 0.07 &             \\  
B  & $1.2\pm0.09$  & $-0.90\pm0.13$ & $0.13\pm0.04$ & 0.029 & 0.17  & $0.83\pm0.13$ & $0.72\pm0.13$ \\
C  & $5.0\pm0.4$   & $-1.02\pm0.16$ & $0.64\pm0.14$ & 0.088 & 0.51  & $0.90\pm0.09$ & $0.82\pm0.09$ \\
D  & $3.4\pm0.2$   & $-0.93\pm0.10$ & $0.20\pm0.07$ & 0.053 & 0.25  & $0.76\pm0.17$ & $0.61\pm0.17$ \\
E  & $6.2\pm0.5$   & $-0.90\pm0.13$ & $0.58\pm0.19$ & 0.164 & 0.52  & $0.80\pm0.20$ & $0.68\pm0.20$ \\
\bottomrule                                                      
\end{tabular}

\caption{Derived properties of regions listed  in
  Table~\ref{tab:regions}. Column $\nu_{\rm SN}$ is the supernova rate, $\alpha$ is the index of the power-law synchrotron component, $\Psi$ is the star-formation rate derived from the free-free component, $\Psi_{\Halpha}$ is the star-formation rate derived from \Halpha\ emission and $\Psi_{24\,\micron}$ is the star-formation rate derived from 24\,\micron\ mid-infrared emission.  Columns  $f_{{\rm th},90\,\unit{GHz}}$
  and $f_{{\rm th},33\,\unit{GHz}}$ are the predicted fractions of
  total flux density that are due to free-free emission at 90 and 33\,GHz
  respectively and are  computed by marginalisation  of the model over
  the full joint  posterior distribution.
  $^*$Since the model fit for region `A'
  is poor, this SFR was derived by direct
  application of Equation~\ref{eq:freefreemodel}.} 

\label{tab:derivedtable}

\end{table*}

\begin{figure}
  \includegraphics[width=0.99\columnwidth]{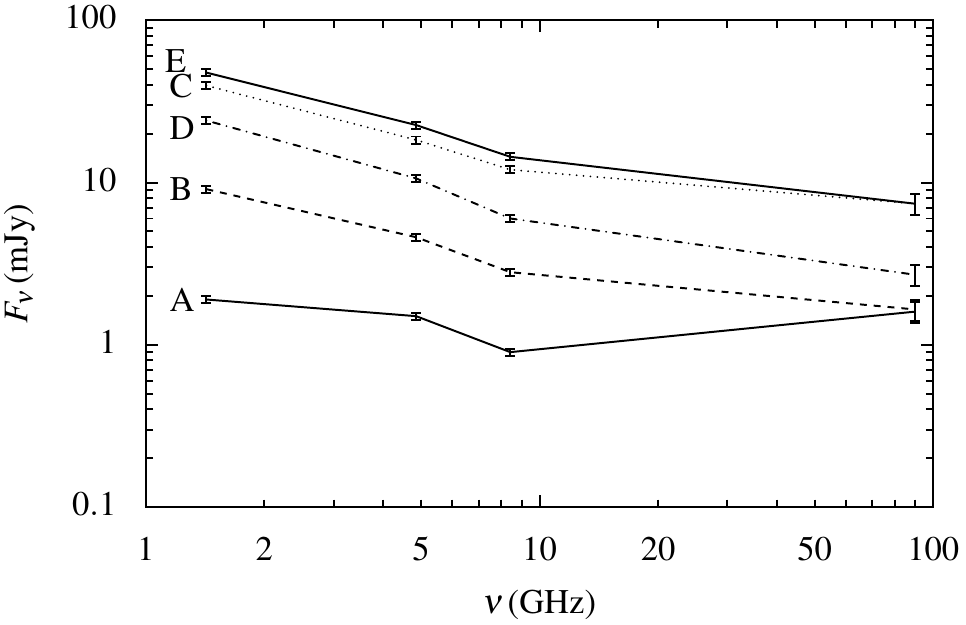}
  \caption{Observed flux densities of the five regions with detections
    at 90\,GHz (right-most points) and at other frequencies with
    matched resolution we consider in this letter. }
  \label{fig:fiveregs}
\end{figure}

\subsection{Models}

We consider two models for the radio spectra. The first model consists
of a simple power-law synchrotron component \cite[calibrated using
relations in ][]{1992ARA&A..30..575C} and a free-free component that
was also modelled as a power law and was calibrated following
\cite{2011ApJ...737...67M}. The second model includes the possibility
of a curvature or break in the synchrotron component by modelling it
as a parabola in log-log space as done by \cite{2010ApJ...710.1462W}.

The synchrotron component can therefore be represented in both models
by:
\begin{align}
  L_{\nu,\rm SN} &= \frac{\nu_{\rm SN}}{1\,\unitp{yr}{-1}}  1.3\times10^{23}\,\unit{W}\,\unitp{Hz}{-1}\, \exp \left(\alpha\xi+\gamma\xi^2\right),
\end{align}
where:
\begin{description}
  \item[$\xi$] is a dimensionless frequency parameter $\xi= \log\left(
      \frac{\nu}{1\,\unit{GHz}}\right)$;
  \item[$\nu_{\rm SN}$] is the rate of supernova explosions;
  \item[$\alpha$] is the linear slope of the synchrotron spectrum (i.e. the spectral index, 
  if the synchrotron spectrum is not curved);
  \item[$\gamma$] is a parametrisation of the curvature of the synchrotron
    spectrum, and is {\it fixed at zero for the simple power-law model}.
\end{description}
The free-free component is given by:
\begin{align}
  L_{\nu,\rm FF}=2.17\times10^{20}\,\unit{W}\,\unitp{Hz}{-1}\,  \frac{\Psi}{1\,\unit{\MSolar}\,\unitp{yr}{-1}}
  \left(\frac{\nu}{1\,\unit{GHz}}\right)^{-0.1}
\label{eq:freefreemodel}
\end{align}
where $\Psi$ is the star-formation rate. This is an update by
\cite{2011ApJ...737...67M} of the \cite{1992ARA&A..30..575C}
relationship using the initial mass function as described by
\cite{2001MNRAS.322..231K} and extending the lower limit of the mass
function to $0.1\,\MSolar$.

\subsection{Analysis}

We performed Bayesian statistical inference \citep[see for
example][]{Jaynes:PTLS,RevModPhys.83.943} of the observed flux
densities for each region and these two models using the {\tt
  radiospec} package described by \cite{2009arXiv0912.2317N}. As
parameters $\nu_{\rm SN}$ and $\Psi$ can be considered scale
parameters we adopt log-flat, non-informative, priors for them. For
parameter $\alpha$ we adopted a flat prior $-1.5<\alpha<-0.5$ and for
parameter $\gamma$ a flat prior $-1<\gamma<0$.

\begin{figure}
\begin{tabular}{c}
  \includegraphics[width=0.99\columnwidth]{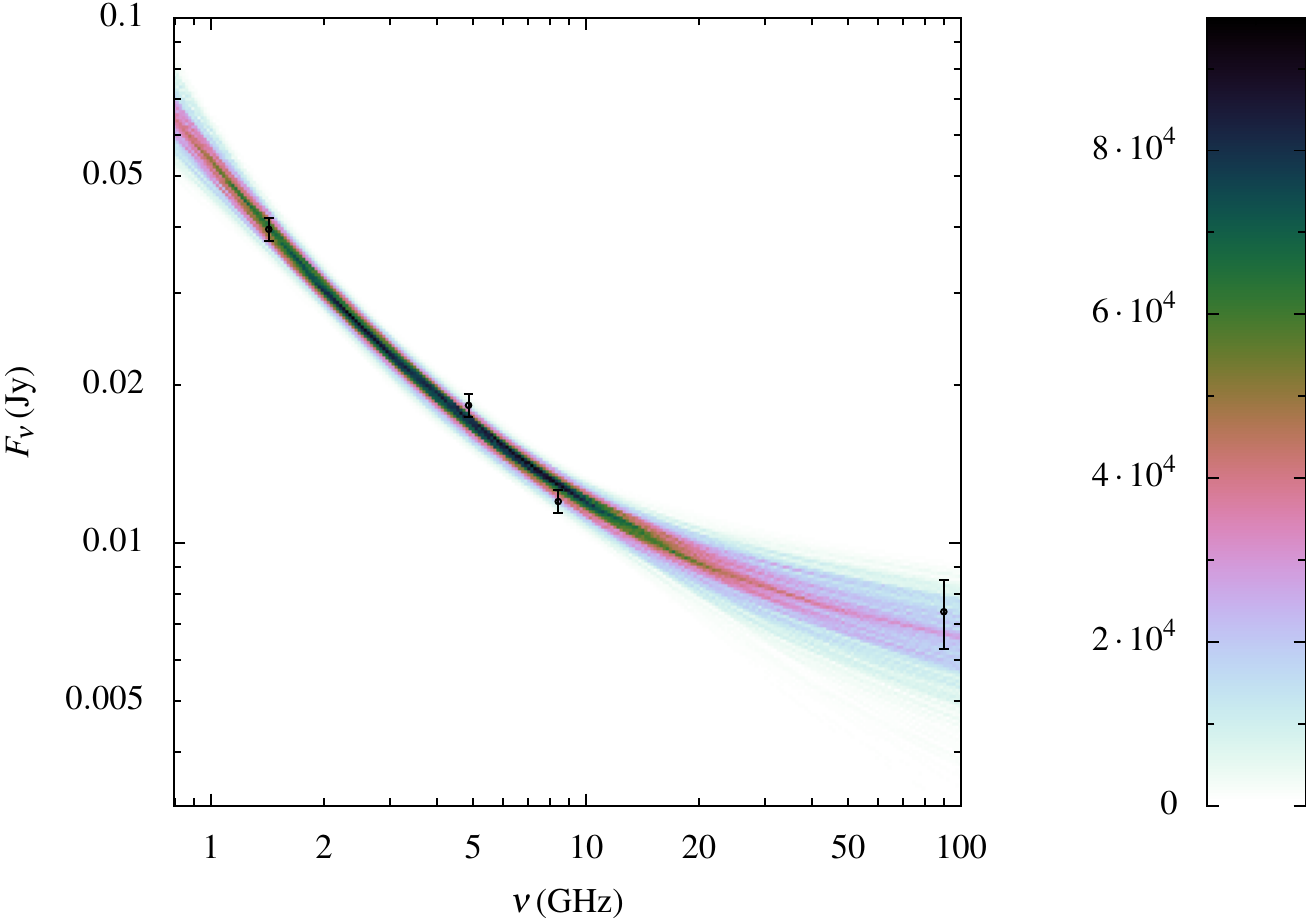} \\
\end{tabular}
\caption{Fan-diagram plot of the fit of power-law synchrotron and
  free-free model for observed radio emission from region `C'. The
  points and error bars are the measured flux densities for this region. The
  colour scale represents the relative probability distribution of
  possible spectra given the power-law synchrotron and free-free
  components model and the inferred posterior distribution of
  parameters. The overall magnitude of the colour scale is set by
  requirement that the integral along each vertical slice of the
  distribution equals the Bayesian evidence (marginal likelihood) for
  this model. }
\label{fig:regioc-fandiagram}
\end{figure}

The computed Bayesian evidence (also known as marginal likelihood) did
not show preference for the curved synchrotron component model in any
region. If we nevertheless adopt the curved synchrotron model
  we can place a limit on the curvature parameter of $\gamma>-0.4$ for
  all of the regions. The analysis would be better able to
  discriminate between these models if more frequency data points were
  available, including at frequencies below 1.4 and around 5\,GHz;
  and, naturally, if the measurement uncertainties (which are
  dominated by calibration uncertainties) were reduced. For the
remainder of analysis we concentrate on the simpler power-law
synchrotron model.

The Bayesian evidence also showed that the quality of the fit to
measured flux densities of region `A' was poor and therefore we have
not used the Bayesian inference to derive physical properties of this
region. Given the inconsistent flux densities at lower radio
frequencies and that region `A' is in a part of our GBT/MUSTANG map
with higher noise, the flux density measurement for region `A' should
be interpreted with caution.

As an illustration of the inference process, we show in
Figure~\ref{fig:regioc-fandiagram} the fan-diagram \citep[see
][]{2009arXiv0912.2317N} of the fit of this model to the brightest
region which is region `C'.  

The Bayesian analysis results in a joint posterior
  probability distribution of all parameters. We have summarised the
  marginalised probability distributions of each parameter in
  Table~\ref{tab:derivedtable} by its mean and standard deviation,
  together with simple estimates of star-formation rates from
  24\,\micron\ and \Halpha\ emission \citep[made using relations given
  by][]{2011ApJ...737...67M}. Such a summary does not however give a
  complete picture of the posterior distribution when it is not
  normally distributed or when there are significant correlations
  between parameters.

\begin{figure}
  \begin{tabular}{c}
     \includegraphics[width=0.99\columnwidth]{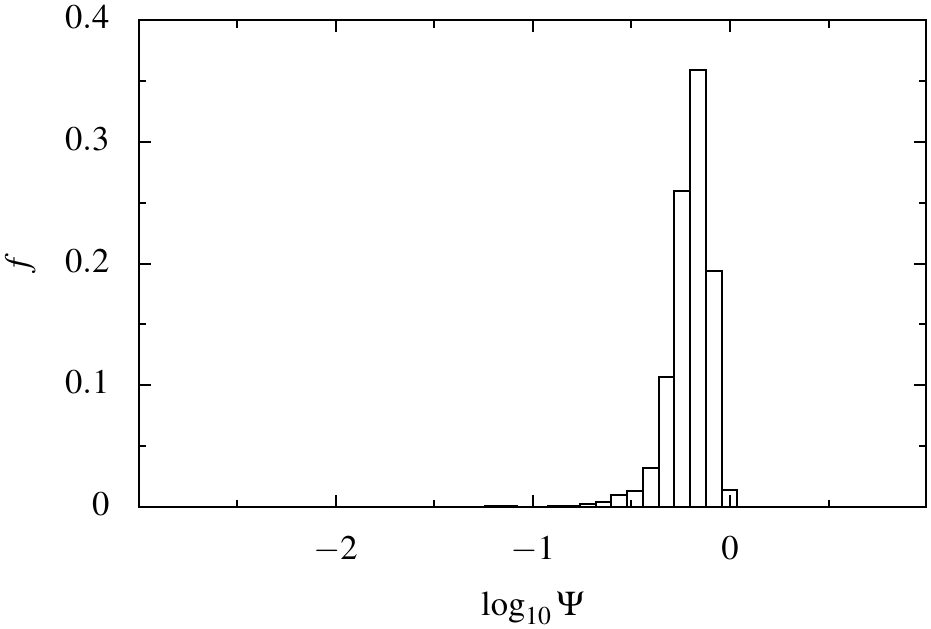}     \\
     \includegraphics[width=0.99\columnwidth]{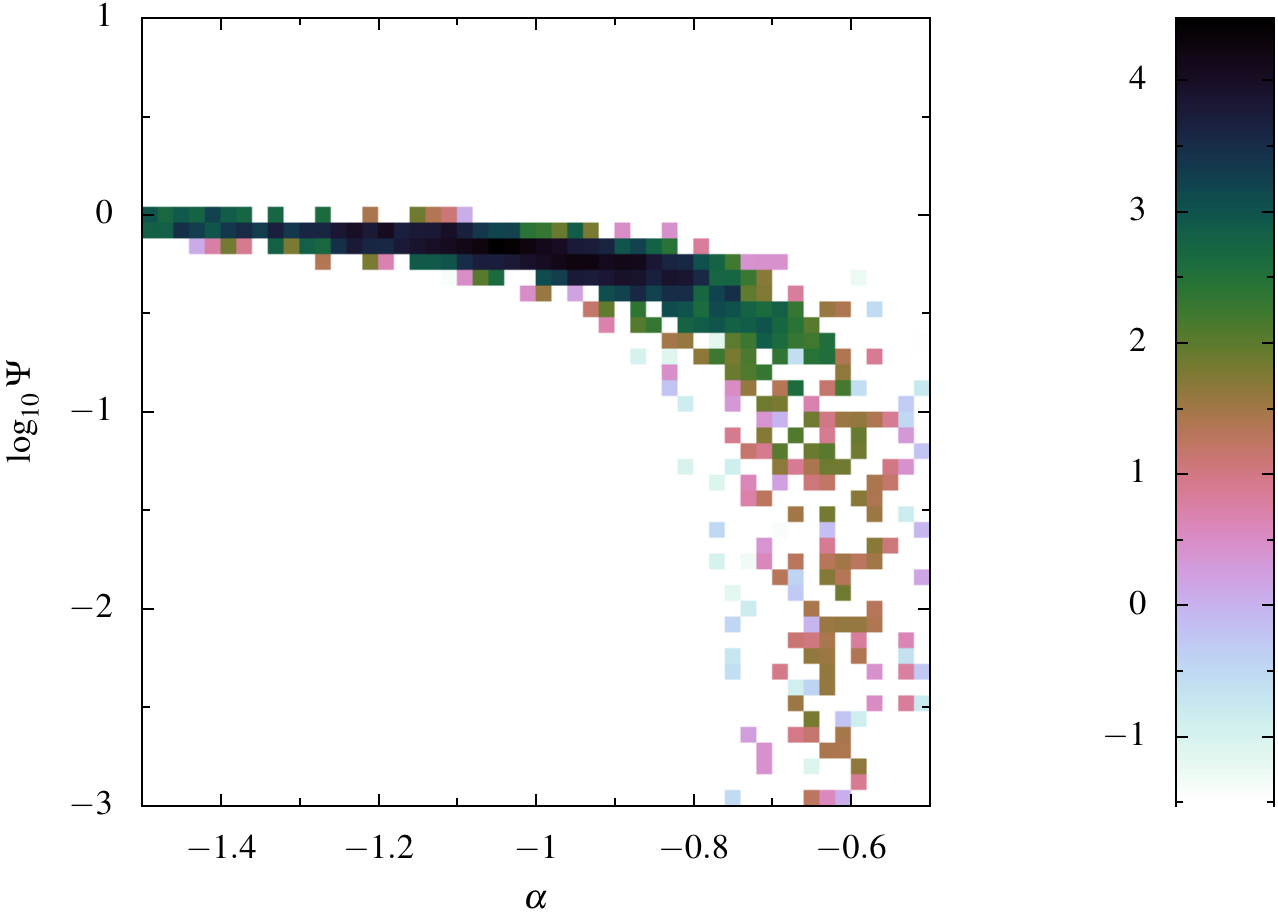} \\
  \end{tabular}
  \caption{Marginal posterior distribution of the $\Psi$ parameter
    (star-formation rate from free-free luminosity, upper panel) and
    joint posterior probability distribution of parameters $\Psi$ and
    $\alpha$ (slope of the synchrotron power law, lower panel) from
    the analysis of region `C'. The colour scale in the lower panel
    represents $\log_{10}$ of the posterior probability which is
    normalised so that its integral over the area of the plot is equal
    to the Bayesian evidence.}
\label{fig:regioc-jointpsialpha}
\end{figure}

  This correlation is most significant in our analysis for parameters
  $\Psi$ and $\alpha$ as illustrated in
  Figure~\ref{fig:regioc-jointpsialpha} for the case of region `C'
  (regions `B'--`E' show similar results). The top panel of this
  figure shows the marginalised posterior probability of $\Psi$ and it
  can be noted that although it has a well defined peak it also shows
  a tail of low values. The reason for this tail of low values is
  explained in the lower panel of this figure which shows there is a
  degeneracy (but at a relatively low probability) between parameters
  $\Psi$ and $\alpha$. Therefore even with measurements at
$90\,\unit{GHz}$, the uncertainty in $\alpha$ still contributes to an
extent to the accuracy with which can $\Psi$ be inferred.

\subsection{Discussion}

It is clear from Figure~\ref{fig:mustang-on-24} that the morphology of
free-free and mid-infrared emission are similar. Quantitatively, the
star-formation rates estimated from free-free emission, $\Psi$, are in
broad agreement with the estimates from 24\,\micron\ emission for
regions `B'--'E', i.e., they are all within one standard
  deviation of the posterior of each other.  It appears therefore
that at these modest star-formation rates and at physical resolutions
of about 1\,kpc there is no deficit in free-free emission
like found in luminous and ultra-luminous infra-red starburst
galaxies by \cite{2010MNRAS.405..887C} and
\cite{2011ApJ...739L..25L} but that instead the measured emission is
consistent with existing models, as also found for nearby galaxies
by \cite{2011MNRAS.416L..99P} and \cite{2011ApJ...737...67M}.
Also shown in Table~\ref{tab:derivedtable} are star-formation
rates estimated from \Halpha\ measurements. These are much (by a
factor of 3 to 6) smaller than the rates derived from 90\,GHz and
24\,\micron\ observations and the primary cause of this is likely to
be internal extinction of the \Halpha\ radiation within M\,66. 

Additionally, the inferred supernova rates $\nu_{\rm SN}$ can also be
compared to the estimates of star-formation rates by multiplying them
by a factor of 86.3 \citep{2011ApJ...737...67M}. We find that for
regions `B', 'C' and 'E' the star-formation rate inferred in this way
underestimates the free-free estimates by an average of 21
  per-cent but this result is statistically weak because the errors
  for each region are large and because the errors are likely to be
  correlated between the regions, i.e. , they are dominated by
  systematic calibration uncertainty which is likely to affect each of
  the regions in a similar way. This discrepancy is significantly
  smaller than the average factor of 2 found by
  \cite{2011ApJ...737...67M} for the case when they (like we) do not
  subtract the local background. This difference may be due to the
  high inclination of M\,66 which means that diffusion of electrons
  within the disc reduces the surface brightness by a smaller factor
  compared to a face-on galaxy, or because the burst of star formation
  in M\,66 is more recent leaving little time for diffusion of
  relativistic electrons. 

We did not consider region `D' in this discussion as it is
  coincident with the nucleus of M\,66 which is classified as
  transition/Seyfert 2 nucleus by \cite{1997ApJS..112..315H} and so
  there is the additional uncertainty of a possible contribution to
  synchrotron and mid-infrared emission from the active galactic
  nucleus.

From the models we also compute and show in
Table~\ref{tab:derivedtable} the fraction of predicted flux density at
90\,GHz and 33\,GHz that is due to free-free emission (the ``thermal
fraction''). We find that the mean thermal fraction at 90\,GHz is
about 83 per cent but is also consistent (due to large
estimated error) with essentially all of the emission being due to the
free-free component. At 33\,GHz we predict a thermal fraction of
around 71 per cent which is consistent with the 79 per cent
that \cite{2011ApJ...737...67M} find when local diffuse background is
not subtracted.

The detection of region `A' at 90\,GHz is at the level of
  four standard errors so it is somewhat statistically uncertain. If
  the true flux density of this region is indeed close to our measured
  value then the spectrum is rising between 8.5 and 90\,GHz which is
  not possible in the models we used in the analysis and consequently
  the Bayesian evidence of both of our models for this region is very
  poor. For this reason we do not show the results of Bayesian
  analysis for this region. A possible explanation of such a rising
  spectrum would be a very compact, high density \HII\ region which
  should be easy to test using high spatial resolution observations at
  90\,GHz using an interferometer.

Finally, we briefly discuss advantages and efficiency of
  bolometer observations at 90\,GHz of nearby galaxies such as those
  presented in this paper.  Our results above show that close to all
  of the flux density at 90\,GHz is due to free-free emission and this
  is one of the primary advantages of such observations, as free-free
  emission is closely related to the number of ionising photons
  emitted by young stars yet unaffected by dust obscuration. Hence
  observations at 90\,GHz can be used to estimate the rate of ionising
  photon absorption by hydrogen atoms without uncertainties associated
  with correction for dust extinction of \Halpha\ measurements. This
  is essentially what the $\left<\Psi\right>$ column in
  Table~\ref{tab:derivedtable} measures (after conversion to
  star-formation rate) and it can be noted that our estimated
  uncertainty is still 20--30 per cent. This is due to the relatively
  large uncertainty we assume for our absolute calibration of our
  observations at 90\,GHz (15 per cent) and uncertainty in removal of
  the small but not well known synchrotron component at this
  frequency.  Experience with bolometer observations at 90\,GHz and
  improvements in processing algorithms should reduce the absolute
  calibration uncertainty in the future while further studies at
  intermediate frequencies should help us better constrain and
  understand the synchrotron component in star-forming galaxies at
  these frequencies.

  Star-formation which can be measured with observations like ours can
  be approximately estimated by combining
  Equation~\ref{eq:freefreemodel} and the noise in our maps which is
  approximately 0.5\,mJy/beam in the best parts. This leads to
  estimate of $1\sigma$ sensitivity to unresolved star formation of
  about $(D/10\,\unit{Mpc})^20.05\,\unit{\MSolar}\,\unitp{yr}{-1}$,
  where $D$ is the distance to the galaxy, while for resolved star
  formation it is about
  $0.22\,\unit{\MSolar}\,\unitp{yr}{-1}\,\unitp{kpc}{-2}$. Our
  non-detection of Messier\,99 is consistent with these sensitives
  when we use the {\it Spitzer\/} 24\,\micron\ map of this galaxy to
  predict star formation distribution. Adopting a distance of 20\,Mpc,
  such an estimate leads to star-formation rate of
  $0.36\,\unit{\MSolar}\,\unitp{yr}{-1}$ in the compact central region
  of this galaxy which is below the $2\sigma$ level expected at this
  distance.

\section{Conclusions}

We present results of a pilot project to observe nearby galaxies with
GBT/MUSTANG to high surface brightness sensitivity. We find:
\begin{enumerate}
\item GBT/MUSTANG can measure the continuum emission at 90\,GHz from
  nearby star-forming galaxies with resolution around
    9\,arcsec and in our case $1\sigma$ sensitivity to unresolved sources that
    corresponds to a star-formation rate of approximately
    $(D/10\,\unit{Mpc})^20.05\,\unit{\MSolar}\,\unitp{yr}{-1}$;
\item These measurements in turn constrain the free-free emission
  associated with ionising radiation from young stars ;
\item For the case of M\,66, the measured free-free emission is
  consistent, to within one standard deviation, with
    star-formation rate estimates from 24\,\micron\ {\it Spitzer}
    observations; the free-free derived star-formation rates are
    higher by an average of 21 per cent than the rates derived from
    low frequency radio data but this is a statistically weak
    discrepancy;
\item Analysis of measurements at L, C, X-band and 90\,GHz do not
  prefer a curved synchrotron specturm as opposed to a simple power
  law synchrotron and free-free components model;
\item The estimate of the thermal fraction at 90\,GHz is close to
  unity; at 33\,GHz it is estimated to be around 71 per cent.
\end{enumerate}

\section*{Acknowledgements}

The authors would like to thank the MUSTANG instrument team from the
University of Pennsylvania, NRAO, Cardiff University, NASA-GSFC, and
NIST for their efforts on the instrument and software that have made
this work possible. We extend particular thanks to Brian Mason, for
his help guiding us through the subtleties of the data reduction
process.

The National Radio Astronomy Observatory is a facility of the National
Science Foundation operated under cooperative agreement by Associated
Universities, Inc.  

We thank the anonymous referee for a prompt referee's report with
suggestions that have substantially improved this paper.  We thank the
SINGS team for making publicly available the comprehensive data on the
galaxies in the SINGS sample.  We would like to acknowledge helpful
comments on some of the early results of this work by K. Johnson and
G.~J. Bendo.

\bibliographystyle{mn2eurl} 
\bibliography{m66letter.bib}

\label{lastpage}

\end{document}